\documentclass{aa}  
\usepackage{graphicx}
\usepackage{array}
\usepackage[varg]{txfonts}
\usepackage{graphicx}
\usepackage{natbib,twoopt}
\usepackage[breaklinks=true]{hyperref} 
\bibpunct{(}{)}{;}{a}{}{,} 
\makeatletter
\newcommandtwoopt{\citeads}[3][][]{\href{http://adsabs.harvard.edu/abs/#3}%
{\def\hyper@linkstart##1##2{}%
\let\hyper@linkend\@empty\citealp[#1][#2]{#3}}}
\newcommandtwoopt{\citepads}[3][][]{\href{http://adsabs.harvard.edu/abs/#3}%
{\def\hyper@linkstart##1##2{}%
\let\hyper@linkend\@empty\citep[#1][#2]{#3}}}
\newcommandtwoopt{\citetads}[3][][]{\href{http://adsabs.harvard.edu/abs/#3}%
{\def\hyper@linkstart##1##2{}%
\let\hyper@linkend\@empty\citet[#1][#2]{#3}}}
\newcommandtwoopt{\citeyearads}[3][][]%
{\href{http://adsabs.harvard.edu/abs/#3}
{\def\hyper@linkstart##1##2{}%
\let\hyper@linkend\@empty\citeyear[#1][#2]{#3}}}
\makeatother
\bibpunct{(}{)}{;}{a}{}{,}

\begin{document}
\title{Colliding interstellar bubbles in the direction of $l=54^{\circ}$}
 
   \author{L. Zychov\'{a}\inst{1}
	\and S. Ehlerov\'{a}\inst{2}}

\institute{Department of Theoretical Physics and Astrophysics, 
Faculty of Science, Masaryk University, 
Kotl\'a\v rsk\'a 2, 611 37 Brno, Czech Republic
         \and
Astronomical Institute, Academy of Sciences of the Czech Republic, 
Bo\v{c}n\'{\i} II 1401, 141 31 Prague 4, Czech Republic}

\date{Received December 4, 2015; accepted August 30, 2016}

 \abstract
{Interstellar bubbles are structures in the interstellar medium with diameters of a few to tens of parsecs. Their progenitors are stellar winds, intense radiation of massive stars, or supernova explosions. Star formation and young stellar objects are commonly associated with these structures.}
{We compare IR observations of bubbles N115, N116 and N117 with atomic, molecular and ionized gas in this region. While determining the dynamical properties of the bubbles, we also look into their ambient environment to understand their formation in a wider context.}
{For finding bubbles in HI (VLA Galactic Plane Survey) and CO data (Galactic Ring Survey), we used their images from Galactic Legacy Infrared Mid-Plane Survey. We manually constructed masks based on the appearance of the bubbles in the IR images and applied it to the HI and CO data. We determined their kinematic distance, size, expansion velocity, mass, original density of the maternal cloud, age and energy input.}
{We identified two systems of bubbles: the first, background system, is formed by large structures G053.9+0.2 and SNR G054.4-0.3 and the infrared bubble N116+117. The second, foreground system, includes the infrared bubble N115 and two large HI bubbles, which we discovered in the HI data. Both systems are independent, lying at different distances, but look similar. They are both formed by two large colliding bubbles with radii around 20--30 pc and ages of a few million years. A younger and smaller ($\sim$4~pc, less than a million years) infrared bubble lies at the position of the collision.}
{We found that both infrared bubbles N115 and N116+117 are associated with the collisions of larger and older bubbles. We propose, that such collisions increase the probability of further star formation, probably by squeezing the interstellar material, suggesting that it is an important mechanism for star formation.}

	
	 \authorrunning{L. Zychov\'{a}}
	 \titlerunning{Colliding interstellar bubbles in the direction of $l=54^{\circ}$}
	
   \maketitle
%

\section{Introduction}
\subsection{Interstellar bubbles}

The interstellar medium (ISM) in the Galaxy shows 
a rich variety of structures, including clumps, filaments and dense sheets.
Some of the structures resemble envelopes and are known as shells or 
bubbles. Their (nearly) ubiquitous presence is the reason behind
the term ``the bubbling galactic disk''  \citep{churchwell2006, deharveng2010, simpson2012},
describing the state of the ISM. Progenitors of these
bubbles, which are best observable in the infrared band, include stellar winds, intense radiation
of massive stars, or supernova explosions. Basically all infrared (IR) bubbles are found near HII regions,
CO clouds or other indicators of the recent or on-going star formation.

The walls delineating the IR bubbles are associated with sites of the new star formation,
i.e. they contain objects younger than the predicted age of the bubble
itself -- many examples can be found in \citep{deharveng2010},
suggesting triggered star formation. 
There are two basic mechanisms proposed to explain such triggering: 
Collect \& Collapse (C\&C) and Radiation Driven Implosion (RDI).
In the first one, the interstellar medium is swept up into a dense
cold layer and this layer gravitationally fragments and then forms
stars. In the second scenario the shock front compresses existing
cold clumps and thus initiates the star formation. 

It is difficult to distinguish between these scenarios 
\citep[see, e.g.][for the discussion of these questions connected
with the IR bubble RCW 120]{walch2015}, 
and there is still the possibility that no real triggering is taking place, just shifting and relocation
of material, which would also form stars without any external driver 
\citep[see][for a discussion about what triggering really
means]{dale2015}.

In this paper, we compare the IR observations of bubbles
N115, N116 and N117 from \citet{churchwell2006} with
atomic (HI), molecular (CO) and ionized (radio continuum) gas 
in this region. We find two independent subsystems, which both
look very much alike: colliding bubbles with the star formation
taking place in the position of the collision. In the following section
we first summarize what is known about the studied bubbles and
then we describe the data and the methods of the analysis. Afterwards,
we will describe each subsystem separately and calculate its properties.
We conclude the paper with a brief discussion.

\subsection{The region around infrared bubbles N115, N116 and N117}

Infrared bubbles N115, N116 and N117 were identified by \citet{churchwell2006}. Radial velocities
associated with these bubbles are $v_{\mathrm{LSR}} = 24.0\ kms^{-1}$ for N115 \citep{watson2010}
and $v_{\mathrm{LSR}} = 42.8\ kms^{-1}$ for N117 \citep{watson2003}\footnote{\citet{watson2010} gives the 
$v_{\mathrm{LSR}} = 18.7\ kms^{-1}$ for N117, but the cited value is not found in the given reference --- as also 
noted by \citet{xu2014}  --- and therefore we disregard it.};
these velocities belong to the (molecular or ionized) gas associated with these bubbles. 

\begin{figure}[h]
\centering
\includegraphics[width=9.5cm]{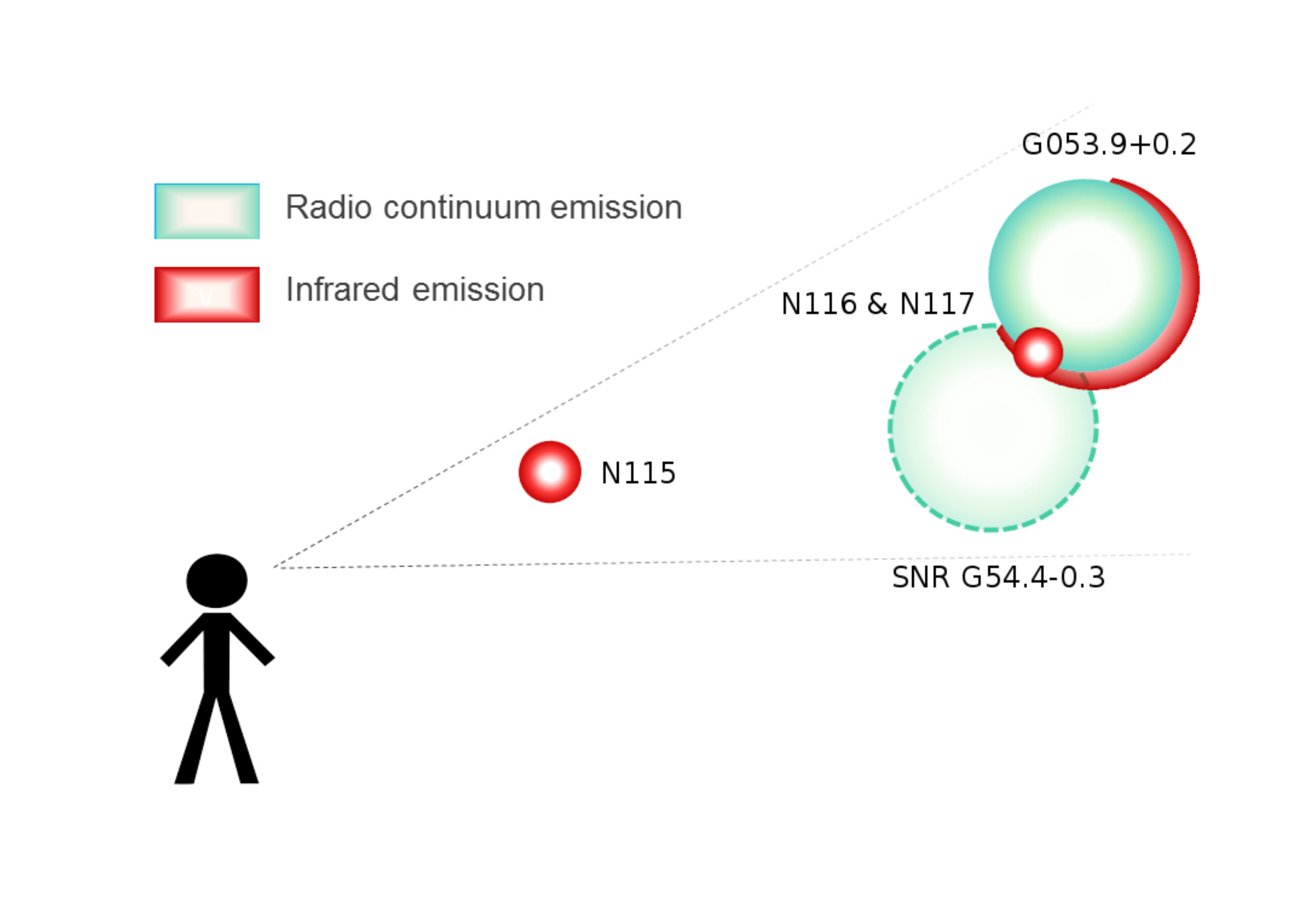}
\caption{A cartoon of the region surrounding G053.9+00.2 according to \citet{xu2014}: 
     N116+117 connected to the radio loop G053.9+0.2 and the independent N115 and SNR G54.4-0.3.} 
\label{before}
\end{figure}

There are two large radio objects in the vicinity of the bubbles: SNR G54.4-0.3 and a radio loop
G053.9+0.2 (see Fig. \ref{background_tot}, the left panel). SNR G54.4-0.3 was studied by \citet{junkes1992} and found to be associated with the
CO gas at around $v_{\mathrm{LSR}} = 40.0\ kms^{-1}$. The loop G053.9+0.2 was studied by \citet{leahy2008}
and was found to be related to an IR shell, which itself was connected to CO emission and HII regions
with velocities around $v_{\mathrm{LSR}} = 40.0\ kms^{-1}$. The CO and the HII regions in the vicinity of the loop G053.9+0.2 were extensively studied by
\citet{xu2014} and the connection with IR bubbles N116 and N117 was explicitly claimed (see Fig. \ref{before} for
their image of the region).

 The pulsar wind nebula, connected with the
SNR054.1+0.3 \citep{green2009}, is projected to the wall of the larger radio loop, but according
to \citet{leahy2008}, is not physically associated with it, as it probably lies at a different
radial velocity ($53\ kms^{-1}$). This pulsar wind nebula is created by the young pulsar PSR J1930+1852 
(the characteristic age of the pulsar is 2900 yr).

\section{Data and methods}

\subsection{Data}

For our study, we use the following surveys:
\begin{itemize}
\item{Spitzer-GLIMPSE survey \citep[Galactic Legacy Infrared Mid-Plane Survey 
Extraordinaire,][]{benjamin2003} with a pixel resolution of $1.2''$.
We use the 8.0 $\mu$m filter.}
\item{VGPS \citep[VLA Galactic Plane Survey,][]{stil2006}
for the HI and radio continuum data. The angular 
resolution is $1'$, the channel width for HI is 1.2 km/s.}
\item{GRS \citep[Galactic Ring Survey,][]{jackson2006} for CO data.
The angular resolution is $46'$, the channel width is 0.2 km/s.}
\end{itemize}

\begin{figure}
\centering
\includegraphics[width=7cm]{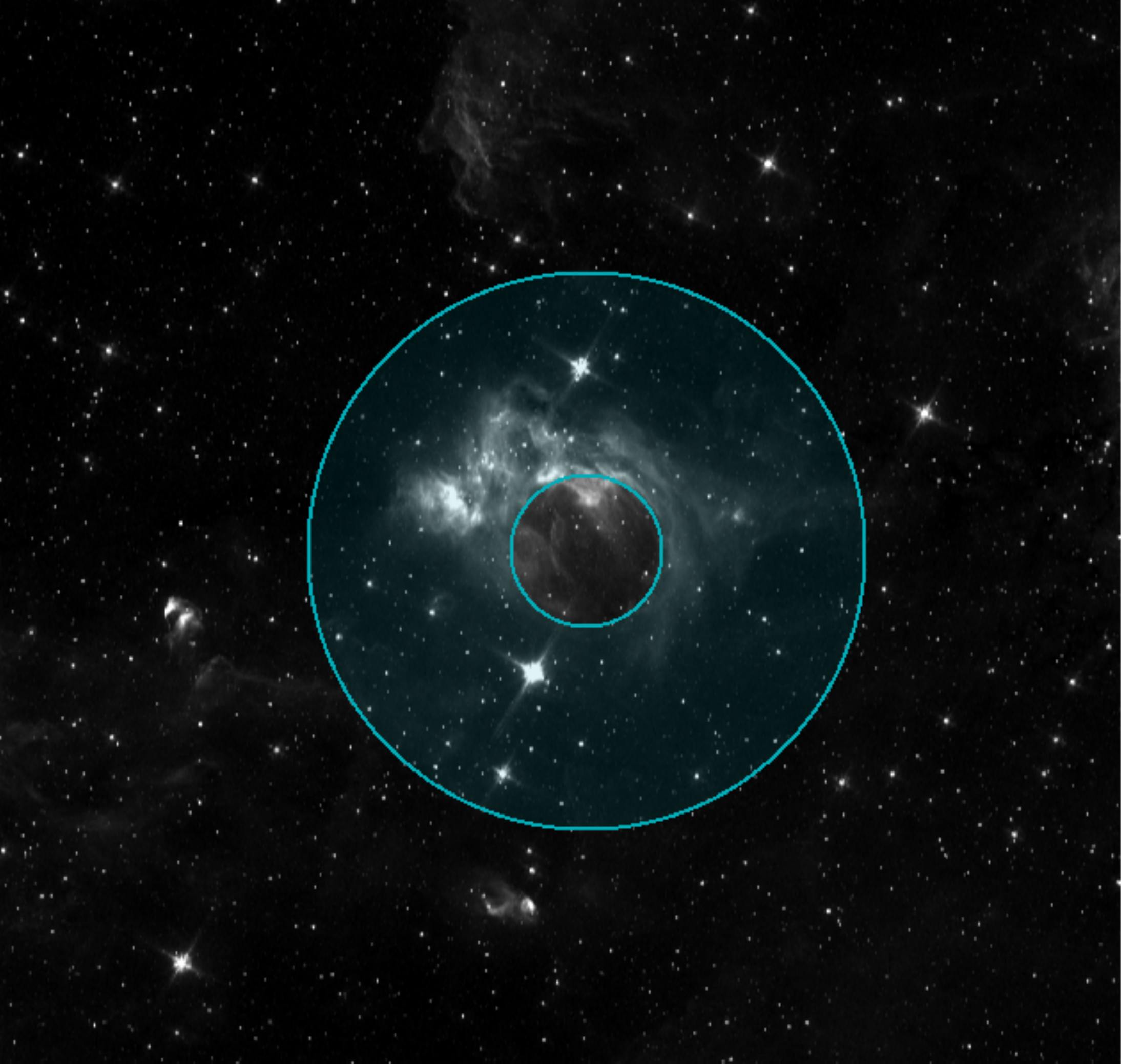}
\caption{The example of a mask for a case of the bubble F (= IR bubble N115).}%
\label{mask}
\end{figure}

We searched the surroundings of known IR bubbles N115, N116 and N117. N116 and N117 are listed separately 
in \citet{churchwell2006} but based on their appearance  and in accordance with HI, CO and radio continuum 
observations we consider them to be one physical entity (see also the discussion in Section 3.1). 
We manually constructed a mask (see the example in Fig.~\ref{mask}) based on the appearance of bubbles
in the IR data (using the ds9 tool -- \citet{ds9}) --- basically a ring --- 
and applied it to the HI and CO data. 
We searched for regions with significant contrast between
the temperature of the ring and temperatures inside and outside 
the ring. We also applied masks, which were 25 percent wider or narrower than the original mask. However, the results did not differ significantly from each other. Such discovered regions were then inspected visually to confirm
the connection between the infrared and HI or CO structures.
Even though there are previous proposed connections between studied IR bubbles and the CO emission
(see the subsection 1.2 in the Introduction), we do not use this a priori knowledge in our search to see, if we will come
to the same conclusions. HI data, to our best knowledge, were not studied in the connection with these
bubbles.

As another source of information about the distribution of the cold phase of the interstellar
medium we use ``BGPS clumps'' from \citet{dunham2011}. These are sources detected in the Bolocam Galactic
Plane Survey (1.1 mm), which are also detected in the $NH_3$ line by the Green Bank Telescope. Diameters of 
these structures lie in the range $33''$ and $5.9'$, but the objects in our studied region are smaller than
$2.1'$, which correspond to linear sizes between 0.3 pc and 3 pc for our assumed distances. These sizes lie
in the interval of sizes given for clumps by \citet{bergin2007}, i.e. objects smaller than clouds and larger
than cores. That is why we adopt the name ``BGPS clumps'' for these tracers of the dense gas.

\subsection{Determination of bubbles properties}
      
Knowing the line-of-sight velocity of a bubble we compute
its kinematic distance using the flat rotation curve 
($R_{\sun} = 8.5\ \mathrm{kpc}$, $V_{\sun} = 220\ \mathrm{kms^{-1}}$).

We derive masses of its atomic and molecular component by using 
formulas from \citet{rohlfs1996}. The neutral hydrogen column density 
$N_\mathrm{HI}$ is derived by:
\begin{equation}
N_{\mathrm{HI}} = 1.8 \times 10^{18} \int{T_{\mathrm{b}}} {} \mathrm{d}v ,	
\label{eq:NHI}
\end{equation}	

where $T_{\mathrm{b}}$ is observed brightness temperature of the HI line. 

The $^{13}$CO column density $N_\mathrm{^{13}CO}$ is derived by:
\begin{equation}
N_{^{13}\mathrm{CO}} = 2.6 \times 10^{14} 
\frac{T_{\mathrm{ex}}}{1-\mathrm{e}^{-T_0/T_{\mathrm{ex}}}} \int{\tau (v)}
\mathrm{d}v ,	
\label{eq:NCO}
\end{equation}	

where $T_{\mathrm{ex}}$ is the excitation temperature, which we assume to 
be $30$ K. $T_0$ for the frequency 110 GHz of the $^{13}$CO ($J = 1-0)$ 
is $5.3$ K. $\tau$ is the optical depth derived by the formula:

\begin{equation}
\tau = -\ln \left[1 - \frac{T_{\mathrm{b}}}{T_0}
\left[(\mathrm{e}^{T_0/T_{\mathrm{ex}}}-1)^{-1} - 
(\mathrm{e}^{T_0/T_{\mathrm{CMB}}}-1)^{-1} \right]^{-1}\right] ,	
\label{eq:tau}
\end{equation}	

where $T_{\mathrm{b}}$ is the observed brightness temperature of the CO line 
and $T_{\mathrm{CMB}}$ is the cosmic mikrowave background temperature of 
$2.7$ K. The GRS survey provides maps of the antenna temperature 
$T_{\mathrm{A}}$, which has to be converted to the brightness temperature:
$T_{\mathrm{b}} = T_{\mathrm{A}}/\eta_{\mathrm{MB}}$, 
where $\eta_{\mathrm{MB}} = 0.48$. 

Using the sizes and column densities we calculate the original density
of the medium, into which the bubble expanded. We give results for
HI and CO separately.

Knowing the radius $r$ of the bubble and its expansion velocity 
$v_{\mathrm{exp}}$ we can compute its age \citep{chevalier1994} $t_{\mathrm{exp}}$ as

\begin{equation}
t_{\mathrm{exp}} = \alpha \frac{r}{v_{\mathrm{exp}}},
\label{eq:age}	
\end{equation}	

where $\alpha = 3/5$ applies to bubbles created by a continuous supply
of energy from an OB association 
and $\alpha = 2/5$ to bubbles created by one-time abrupt supernova explosion.	
	
We estimate the total energy input 
$E_{\mathrm{tot}}$ to create the bubble using the Chevalier's formula:
\begin{equation}
\frac{E_{\mathrm{tot}}}{\mathrm{erg}} = 5.3 \times 10^{43}\left(\frac{n}{\mathrm{cm^{-3}}}\right)^{1.12} \left( \frac{r}{\mathrm{pc}}\right)^{3.12} \left(\frac{v_{\mathrm{exp}}}{\mathrm{km/s}}\right)^{1.40},
\label{eq:etot}		
\end{equation}	
where $n$ is the derived density in the vicinity of the bubble.

We compare our estimates of energies and ages with stellar models 
\citep{schaller1992} and try to derive the masses of the progenitor stars.

\section{Results}

\def\arraystretch{1}
\newcolumntype{M}{>\tiny l} 
\setlength\tabcolsep{1ex}
	
\begin{table*}
\caption{Measured properties of the studied bubbles: positions, radial velocities and diameters, together
with wavelengths, where the bubble is nicely seen or where it was previously detected and described. 'B'
and 'F' stands for the background or foreground system.}
\label{table_names}      
\centering  
\begin{tabular}{l r r r r l l l}   
\hline       
Name &  $l$ [deg] & $b$ [deg] & $v_{r}$ [km/s] & $d$ [arcmin] & Prominent data & System & Alternative name\\
\hline
Bubble A & 53.885	& +0.180	& 41.4	& 39	& radio continuum & B    & Leahy's large-scale bubble \\
Bubble B & 54.498	& -0.303	& 40.0	& 45    & radio continuum & B         	& SNR G54.4-00.3 \\
Bubble C & 54.089	& -0.087	& 40.0	& 5	& IR emission & B         	& N116 \& N117 \\
Bubble D & 53.821	& +0.302	& 25.0	& 78	& HI line & F         	& - \\
Bubble E & 53.257	& -0.419	& 24.9	& 54	& HI line & F         	& - \\
Bubble F & 53.556	& -0.014	& 24.1	& 13	& IR emission & F         	& N115 \\
\hline                  
\end{tabular}
\end{table*}

Based on analysis of both the HI and CO data we find, in agreement with previous observations, 
that infrared bubbles N115 and N116+117 lie at different radial velocities and thus at different distances
from us. 
The bubble N116+117, which has the radial velocity around $v_{\mathrm{lsr}} \simeq 40\ kms^{-1}$, 
was assumed to be connected to a larger radio continuum bubble \citep{xu2014}. We confirm
this and also believe, that it is in fact connected to both radio bubbles seen
in this direction. We also find that the bubble N115 ($v_{\mathrm{lsr}} \simeq 24\ kms^{-1}$)
has two larger accompanying bubbles best visible in the HI gas.

The radial velocity of $40\ \mathrm{kms^{-1}}$ is close to the terminal velocity
in the direction of $l=\ 54^{\circ}$ and therefore we place the N116+117 system
at the tangential point corresponding to the distance of 5 kpc.
The velocity $24\ \mathrm{kms^{-1}}$ has both the near and far kinematic
distances. In accordance with most other authors (and also for reasons
given below) we adopt the near distance as more appropriate. Therefore
we call the system with the radial velocity of $24\ \mathrm{kms^{-1}}$, 
containing the infrared bubble N115, as the foreground system, and the
system with the radial velocity of $40\ \mathrm{kms^{-1}}$ as the background
system.

Table \ref{table_names} gives an overview of the bubbles identified in the area, their
positions and alternate names. Table \ref{table_properties} summarizes their dimensions,
derived masses and densities (Eqs. \ref{eq:NHI}-\ref{eq:tau}), and energies involved in their creation
(Eq. \ref{eq:etot})

\def\arraystretch{1}
\newcolumntype{M}{>\tiny l} 
\setlength\tabcolsep{1ex}
	
\begin{table*}
\caption{Properties of identified bubbles. $R$ is the mean radius, $v_{\mathrm{exp}}$ is the expansion velocity, 
$M_{\mathrm{HI}}$ is the mass of the buble in HI, $M_{\mathrm{H_2}}$ is the molecular mass from, $n$ is the density 
of the medium into which the bubble was expanding (derived from HI and CO), $E$ is the energy needed to create
the bubble. The bubble B is the special case: its expansion velocity does not come from our measurements
and its age is likely to be higher (see the Section 3.1.1).}               
\label{table_properties}      
\centering  
\begin{tabular}{l r r r r r r r r}   
\hline       
Bubble & $R$ & $v_{\mathrm{exp}}$ & $M_{\mathrm{HI}}$ & $M_{\mathrm{H_2}}$ & $n_{\mathrm{HI}}$ & $n_{\mathrm{H_2}}$ & $E$ & age \\
~ & [pc]& [$\mbox{kms}^{-1}$]  & [M$_\odot$] & [M$_\odot$] & [cm$^{-3}$] & [cm$^{-3}$] & [$10^{50}\mbox{erg}$] & [Myr] \\
\hline
A & 28 & 7        & 2.4$\cdot10^{4}$    & 1.4$\cdot10^{5}$   & 13    & 31              & 4.0  & 2.7 \\
B & 32 & 50 $^{*}$ & 1.2$\cdot10^{4}$    & 1.4$\cdot10^{5}$   & 3.4   & 19              & 26.0 & 0.3$^{*}$ \\
C & 4  & 4        & 360                & 3.4$\cdot10^{4}$   & 160   & 7.5$\cdot10^{3}$ & $>$ 0.028 & 0.5 \\
D & 23 & 5        & 2.4$\cdot10^{3}$    & 5.6$\cdot10^{4}$   & 1.8   & 20              & 0.19 & 3.6 \\
E & 17 & 6        & 2.1$\cdot10^{3}$    & 3.0$\cdot10^{4}$   & 3.9   & 26              & 0.21 & 2.3 \\
F & 4  & 4        & 130                & 1.5$\cdot10^{4}$   & 23    & 1.3$\cdot10^{3}$ & $>$ 0.008 & 0.6 \\
\hline                  
\end{tabular}
\end{table*}

\subsection{Background system}

\begin{figure*}
\centering
\includegraphics[height=0.9\textheight]{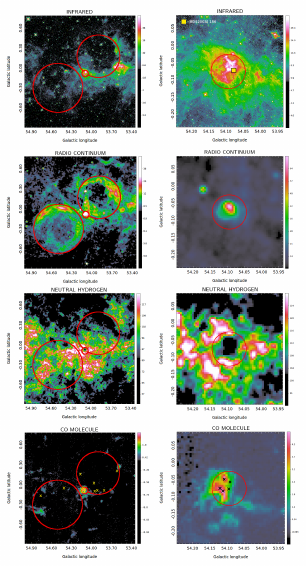}   
\caption{The multiwavelength image of the background system (left: the total system, right: the small central region. 
Positions of bubbles A (upper right), B (lower left) and C (small in the middle) are overlaid. 
The HI is summed over the velocity interval: $38.9\ kms^{-s} < v_{LSR} < 42.2\ kms^{-1}$ (left) and
 $39.8\ kms^{-s} < v_{LSR} < 42.2\ kms^{-1}$ (right), the CO over the interval: $38.6\ kms^{-s} < v_{LSR} < 41.6\ kms^{-1}$ (left)
and $38.4\ kms^{-s} < v_{LSR} < 41.8\ kms^{-1}$ (right).
Crosses in the CO pictures belong to the BGPS clumps found at the corresponding velocity intervals.}%
\label{background_tot}
\end{figure*}

\begin{figure}
\centering
\includegraphics[width=9cm]{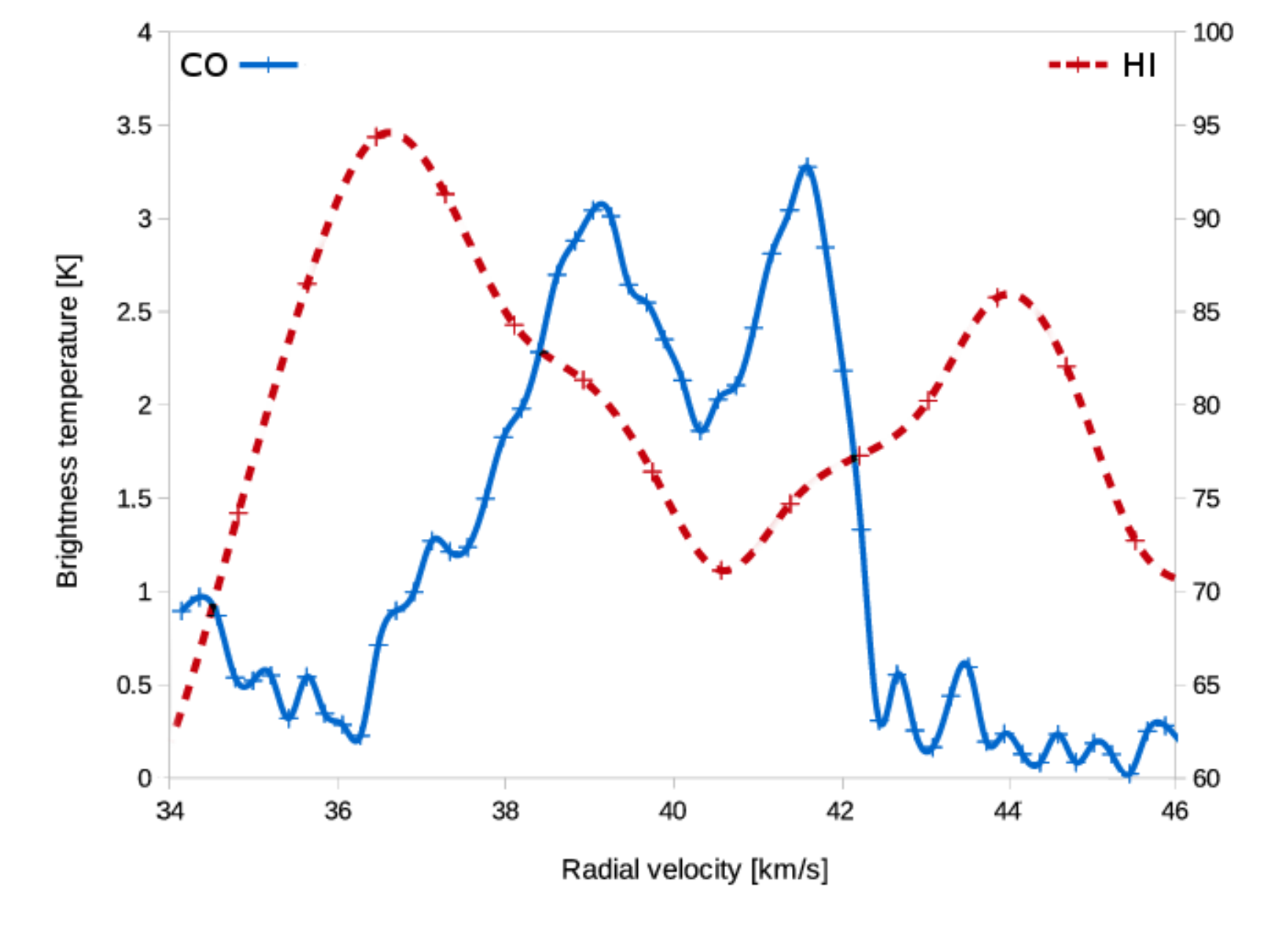}
\caption{The CO (a solid blue line, the left y-axis)  and HI (a dashed red line, the right y-axis) spectrum 
of the bubble C (the background system). The HI spectrum is taken at the position 
($54.087^{\circ}$, $-0.068^{\circ}$), the CO spectrum at the position
($54.123^{\circ}$, $-0.103^{\circ}$).}
\label{spectraC}
\end{figure}

The background system, centered around the IR bubble N116+117 at the radial
velocity of $\sim 40\ \mathrm{kms^{-1}}$, lies at the distance of approximately
5 kpc. For the purposes of the description we will name the larger bubbles of this
system as A and B, and the smaller bubble (which is coincident with the IR bubble
N116+117) as C. 
The diameter of the small bubble C in Figures (and its typical size in all studied
spectral regions) is 5', diameters of A and B are 39' and 45', respectively.  
In linear dimensions it corresponds to 7 pc (C) and 56 or 65 pc (A and B).

In the infrared emission ($8\mu {\mathrm{m}}$) the small bubble C is described by
\citet{churchwell2006} as two bipolar bubbles, one incomplete (N116) and one closed (N117),
see Fig. \ref{background_tot}, the right part. Also the larger bubble A is observable in this spectral region 
as an incomplete ring surrounding a region devoid
of any extended emission (see Fig. \ref{background_tot}, the upper panel the left part; the brightest emission
in the wall of A rightward from C does not belong to this system, it is N115 and belongs to 
the foreground system). The bubble B has no clear counterpart in the IR
region. 

All three bubbles are visible in the radio continuum image. C is the brightest one,
with the radio continuum filling the interior of the infrared structure. A is visible
as the closed irregular ring. B forms the incomplete but well defined spherical ring 
(Fig. \ref{background_tot}, the second panel, the left part).

The bubble best visible in the HI emission is small C. It is a small hole, which is
filled by the radio continuum emission (Fig. \ref{background_tot}, the right half). The hole is surrounded by a fragmented
denser wall. If the structure expands at all, its expansion velocity is small, around
4 $kms^{-1}$. Bubbles A and B are not visible, or at least not visible with a sufficient
credibility. The region is complex and full of arcs, loops and holes, and therefore
one can easily see things which are not there. However, there is a suspicious coincidence
of the HI deficiency and the radio continuum emission at the upper-right part of the
wall in the bubble A and the HI emission just next to this
deficiency, corresponding to the IR emission (Fig. \ref{background_tot}, upper three panels in the left). 
We are unable to measure the expansion velocity of the bubble B, therefore we take
it from \citet{park2013}, and we estimate the expansion velocity of A based on the
mentioned structure. 

The brightest CO emission at around $\sim 40\ kms^{-1}$ lies very near to the bubble C.
The bubble itself does not lie in the centre of the emission but at its edge (Fig. \ref{background_tot}).
Interestingly looking CO arcs protrude from this bright region (they merge at the bubble C).
They have the general direction of the radio continuum emission. Walls of all three bubbles
contain cold molecular clumps (BGPS clumps = crosses in
the lowest panels in Fig. \ref{background_tot}). Fig. \ref{spectraC}
shows the spectrum through the centre of the bubble C (HI) and through the brightest
CO emission inside the bubble (CO). HI expansion velocity is larger than the CO expansion
velocity, which we mostly ascribe to the off-centre position of the CO spectrum.

On the whole it looks like we observe two colliding bubbles. These bubbles (A and B) are
seen as rings of the warm ionized gas. The bubbles probably interact with the neutral gas around them, 
as seen in the distribution of cold molecular gas (in walls) and also partly in the IR emission of the 
heated dust and the HI emission of the atomic gas. Linear radii of these bubbles are around
$30\ pc$. In the region, where these two bubbles collide, there is another, much smaller bubble, 
which is compact, with walls made from the neutral gas and partly traced also by the molecular
gas. This small bubble (C) is filled with the ionized gas.

\subsubsection{Properties of bubbles A, B and C}

Using Eqs. \ref{eq:NHI}-\ref{eq:etot} we calculate masses of individual bubbles (around $10^5M_{\odot}$ for 
A and B and about a third for C), energies needed to create the structures
(using the density $n_{\mathrm{HI}}$ derived from the HI measurements, which means, that this number
is a lower limit for the energy, especially for the small bubble C where the volume filling factor of
the dense molecular gas is quite high) and their ages (2 Myr for A and 0.5 Myr for C). We do not measure the expansion
velocity of B; if we take the value from \citet{park2013} we get the age of 0.3 Myr. 
This age is too young for the bubble B to be able to create the small bubble C by the collision with the
bubble A. However, from the \citet{junkes1992b} and \citet{park2013} it seems plausible, 
that this supernova evolves inside the stellar wind bubble (and still resides inside it) and therefore
the older stellar bubble is the one, which collides with the bubble A.
 
All calculated values can be found in the Table \ref{table_properties}.
Energies involved in the creation of bubbles A and B are quite high
($\sim 4 \times 10^{50} \mbox{erg}$ for A and $\sim 2.6 \times 10^{51} \mbox{erg}$ for B).
This is not surprising for the bubble B, which is the supernova remnant, but even the bubble
A is quite energetic. It could be an old supernova remnant \citep{leahy2008}, but the
needed energy could be supplied by stars, but either a very massive star or several less
massive would be needed. The bubble C is much less energetic and it can be easily
powered by the stellar wind of a star, which is able to ionize its surrounding.

\subsection{Foreground system}

\begin{figure*}
\centering
\includegraphics[height=0.9\textheight]{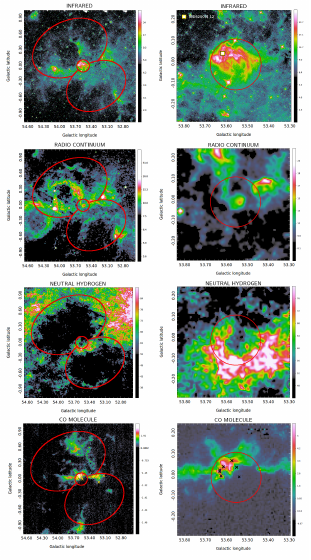}
\caption{The multiwavelength image of the foreground system (left: the total system, right: the small central region). 
Positions of bubbles D (upper left), E (lower right) and F (small in the middle) are overlaid.
The HI is summed over the velocity interval: $23.3\ kms^{-s} < v_{LSR} < 24.9\ kms^{-1}$ (left) and
 $23.3\ kms^{-s} < v_{LSR} < 24.1\ kms^{-1}$ (right), the CO over the interval:
$23.8\ kms^{-s} < v_{LSR} < 25.0\ kms^{-1}$ and $23.9\ kms^{-s} < v_{LSR} < 24.8\ kms^{-1}$ .
Crosses in the CO pictures belong to the BGPS clumps found at the corresponding velocity intervals.}%
\label{foreground_tot}
\end{figure*}

\begin{figure}
\centering
\includegraphics[width=9cm]{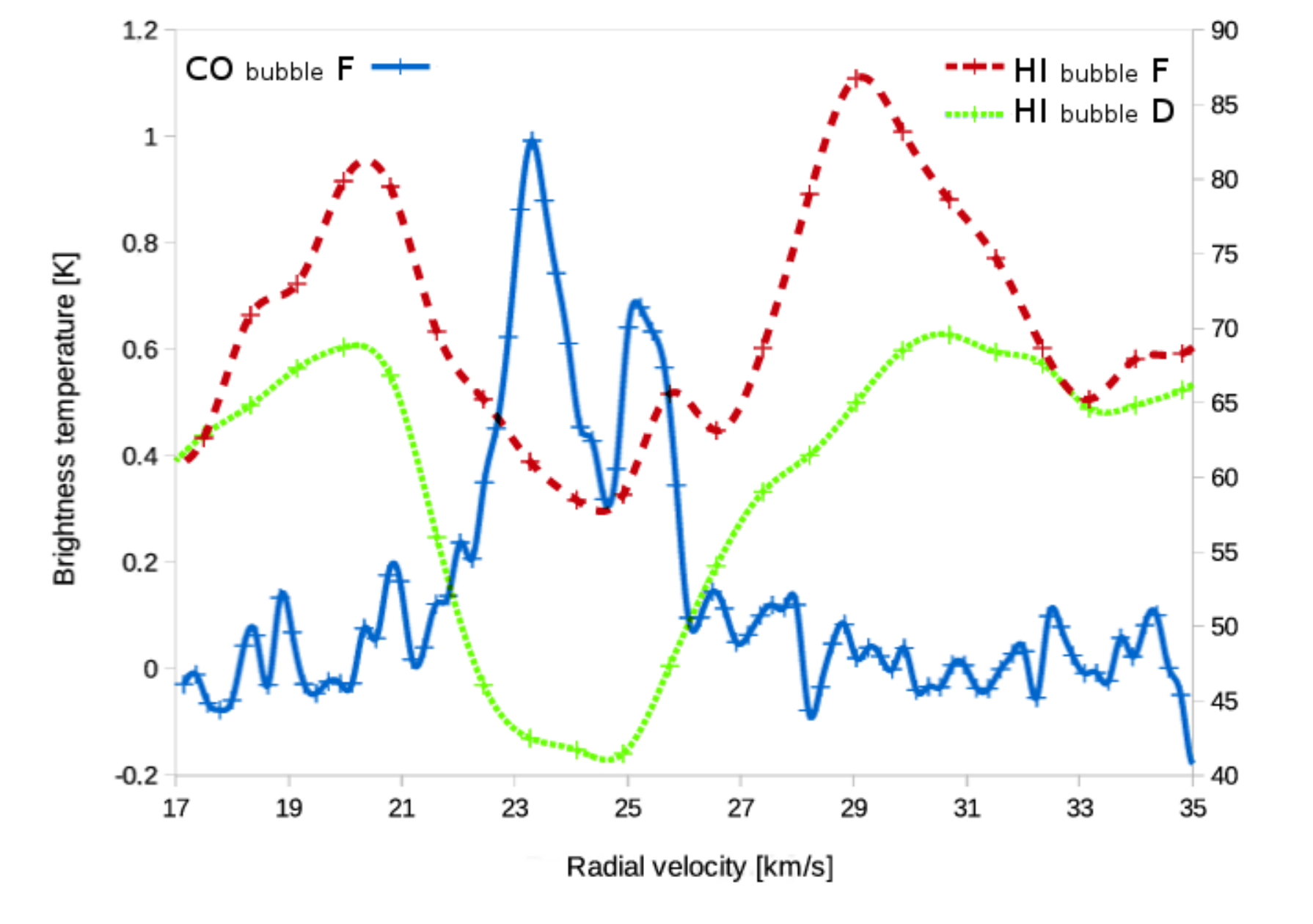}
\caption{The CO (a solid blue line, the left y-axis) and HI (a dashed red line, the right y-axis) spectrum 
of the bubble F (the foreground system). 
The HI spectrum is taken at the position ($53.573^{\circ}$, $-0.059^{\circ}$), the CO spectrum at the position
($53.558^{\circ}$, $0.003^{\circ}$). The HI spectrum of the bubble D (a dotted green line) is overlaid, its position
is ($53.56^{\circ}$, $-0.014^{\circ}$).}
\label{spectraF}
\end{figure}	

\begin{figure}[h]
\centering
\includegraphics[width=9cm]{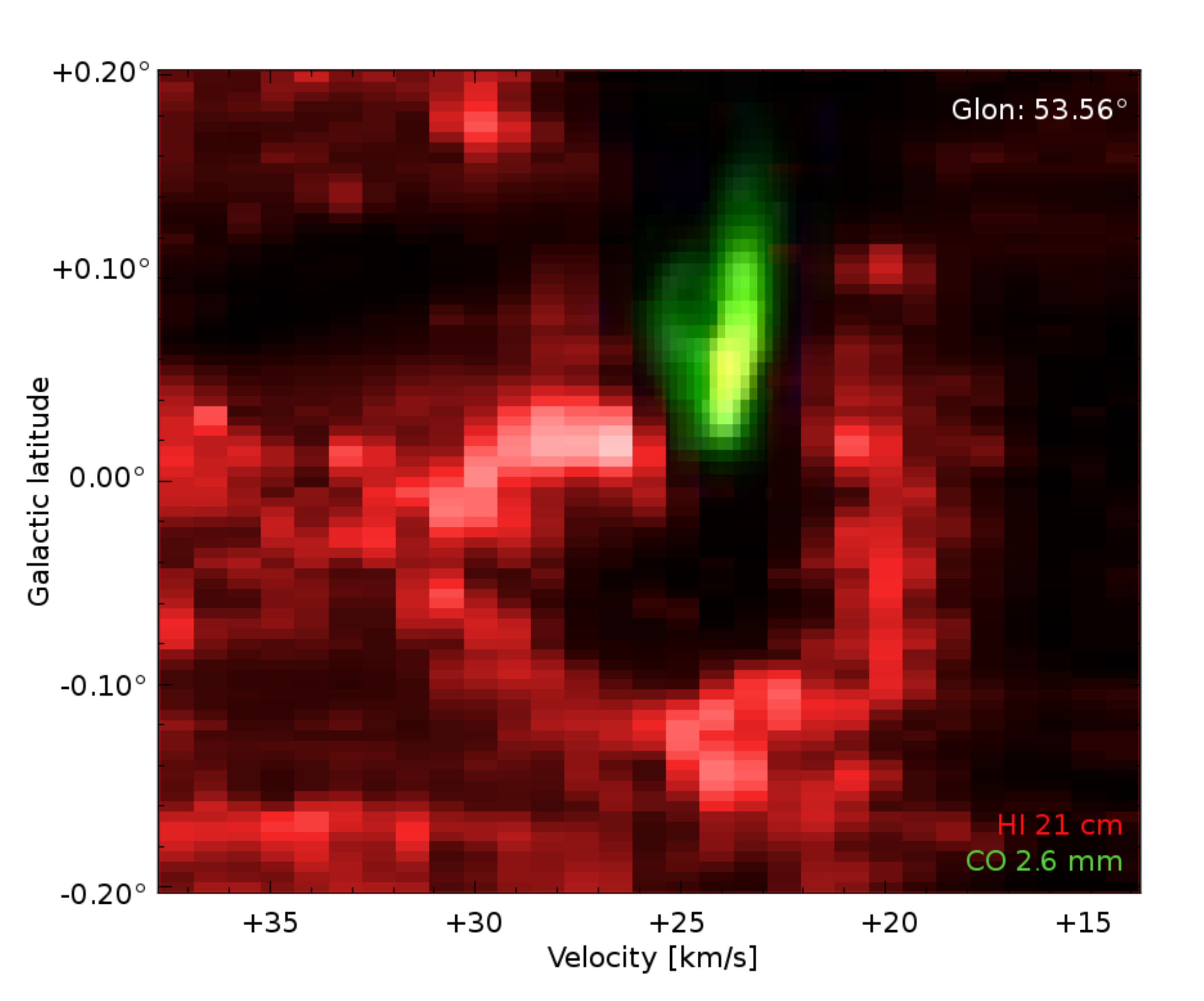}
\caption{The latitude-velocity diagram of the bubble F (the foreground system) at $l=53.56^{\circ}$. 
The red color corresponds to the HI emission, the green color to CO.}
\label{bvF}
\end{figure}

The foreground system, centered around the IR bubble N115 at the radial
velocity of $\sim 24\ \mathrm{kms^{-1}}$, lies at the distance of approximately
2.1 kpc. 
Looking at the 3rd panels of Fig. \ref{foreground_tot} we see, that it contains not
only the small central bubble, but also two larger ones.
We name the larger bubbles of this system as D and E, and the smaller bubble 
(which is coincident with the IR bubble N115) as F. 
The diameter of the small bubble F in Figures (and its typical size in all studied
spectral regions) is 13', diameters of D and E (more precisely, geometrical means of their
axes) are $1.3^{\circ}$ and $0.9^{\circ}$, respectively. 
For the near kinematic distance, which is our adopted distance, these values correspond
to 8 pc (F), 46 pc (D) and 34 pc (E). At the far kinematic distance (8 kpc) sizes of the
larger bubbles D and E would be around 140 pc. Such values are comparable to the thickness
of the HI disk, but we see (Fig. \ref{foreground_tot}, left) that the bubbles are both located inside the disk and
are not influenced by the density gradient in the vertical direction. That is
the other reason, apart from wanting to be on a safe, lower energetics, side, why we
prefer the near kinematic distance.

In the infrared emission ($8\mu {\mathrm{m}}$) the only clearly visible
object is the small bubble F (the IR bubble N115). \citet{churchwell2006} describes it as a closed ring 
(see Fig. \ref{foreground_tot}, the upper panel at the right side). A very faint IR emission might be connected to the 
right-lower wall of the bubble D.

The bubble F is also seen in the radio continuum image. Not very bright emission
comes from the interior of the bubble.  
There might be small and faint regions of the ionized gas along the wall of the bubble D
and also along the upper part of the bubble E (Fig. \ref{foreground_tot}). 

An object at coordinates $l=53.19^{\circ},b=0.16^{\circ}$, visible both in the infrared and radio continuum data, is another IR bubble, MWP1G053179+001558 \citep{simpson2012}. According to
\citet{bronfman1996} and \citet{anderson2009} it lies at the radial velocity 
$6.6kms^{-1} \leq v_{lsr} \leq 8.3kms^{-1}$ and therefore does not belong to our foreground
system (and also not to the background system).

The whole foreground system is well visible in the HI emission (see Fig. \ref{foreground_tot}, the 3rd panel, 
for the channel map at around $v_{lsr} \simeq \ 24kms^{-1}$ and Fig. \ref{spectraF} for the spectrum
through the center of the bubble F). The bubble D is a hole in the HI distribution
in this region, i.e. it shows only the region devoid of HI but no prominent walls
except where it interacts with E (and where F is located). The bubble E
shows, on the contrary, a nice partial wall, or arc, in its lower part. 

The brightest CO emission is connected to the small bubble F (see Fig. \ref{foreground_tot}, and also Fig. \ref{bvF}, 
which is the latitude-velocity diagram of the bubble 
and which shows how the HI and CO together surround the hot interior of the bubble and supplement each other). 
Molecular gas in bubbles D and E is located preferentially at the inner 
edge of the HI structures (it is evident especially in D). One possible
scenario is, that the shock wave, which created D or E, was strong enough
to push the gas, but not strong enough to completely dissolve it (see \citet{ehlerova2016}
for the analysis, how do properties of CO clouds change with the position of clouds
inside and outside HI shells). 
Bubbles F and E contain cold molecular clumps (BGPS clumps = crosses in 
the lowest panels in Fig. \ref{foreground_tot}),
mostly along the wall, where bubbles connect.

The picture which we see in this foreground system resembles the background one:
two larger, and therefore probably older, colliding bubbles, with the region of the very 
recent star formation located in the interacting zone. This zone is the center of activity 
of the system and is best observable in all studied wavelengths.
Unlike in the previously described case, the interacting
bubbles seem to be less energetic, as they are not connected to a substantial degree
with the ionized gas. This is confirmed by the estimates (Table \ref{table_properties}). Ages of 
interacting bubbles A, B, D and E are similar, but D and E are an order of magnitude less 
energetic than A.

\subsection{Progenitors of bubbles}

\begin{figure}
\centering
\includegraphics[width=9.5cm]{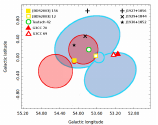}
\caption{A schematic picture of the two systems, background (smaller red one) and foreground
(larger blue one) with young clusters and pulsars in the area.}
\label{bubbles_clusters}
\end{figure}

\def\arraystretch{1}
\newcolumntype{M}{>\tiny l} 
\setlength\tabcolsep{1ex}
	
\begin{table*}
 \caption{Young clusters in the studied area.}             
 \label{table_clusters}      
 \centering     
 \begin{tabular}{M M M M M M}   
 \hline       
Name & $l$ [deg] & $b$ [deg] & $v_{lsr}$ [km/s] & age [Myr] & Type \\
\hline
G3CC 69       &  53.147 &  0.071 & 22.0 & ~ & embedded \\
G3CC 70       &  53.237 &  0.056 & 23.9 & ~ & partially emb. \\
BDS2003 12  &  53.622 &  0.039 & 23.7	& ~ & embedded \\
Teutsch 42    &  53.771 &  0.164 & 37.7 & 3 & exposed, assoc. \\
BDS2003 156 &  54.084	& -0.069 & 39.9 & ~ & embedded \\
\hline                  
\end{tabular}
\end{table*}

One (B) of six studied bubbles is a supernova remnant, one (A) might also be one
and the rest are created by stellar winds. 

The underlying picture for both systems, the background and the foreground, needs the 
coeval and cospatial existence of two or more massive stars, which is indicative of the 
presence of the young cluster.

From \citet{morales2013} we have selected young ($< 100$ Myr) clusters in the area: 
Teutsch 42, [BDS2003] 12 and 156, G3CC 69 and 70. Their positions are shown in the Fig. \ref{bubbles_clusters} 
(together with the schematic positions of bubbles)
and given in Table \ref{table_clusters}. From the
spatial distribution it looks like that [BDS2003] 12 is connected to the bubble F (IR bubble
N115), [BDS2003] 156 to the bubble C (N116+117), Teutsch 42 to either bubble A or D.
G3CC 69 and 70 might be connected to the infrared bubble MWP1G053179+001558 (which
does not belong to any of our studied systems). Radial velocities given in \citet{morales2013}
and cited in Table \ref{table_clusters} belong to the gas associated with the clusters.

[BDS2003] 156 and 12 are both young clusters still embedded in the gas and dust
envelopes \citep{morales2013}.  Their radial velocities correspond to the radial
velocities of the bubbles, which the clusters seem to lie in ([BDS2003] 156 in C = N116+117
and [BDS2003] 12 in F = N 115). The first connection is also noted by \citet{morales2013}.
The unknown but young age of the cluster and its state of being
still embedded in the parental gas satisfies our idea of the star formation 
recently triggered by the colliding bubbles. The newly formed stars then
create and heat the bubbles around them (C and F). 

Clusters G3CC 69 and 70, also young and embedded, lie in the projected vicinity of 
the infrared bubble MWP1G053179+001558. However, the assumed radial velocity
of this bubble is around 7 $kms^{-1}$ (see Section 3.2 for references), and they do not
seem to be physically connected, as the gas around the clusters has the radial
velocity of $\sim 23\ kms^{-1}$. This velocity is the same as the velocity of
the foreground system and therefore if there is any connection, it might be
to this system. Still, they are too young to be progenitors of the larger
bubbles of the foreground system.

The last cluster in this area (Teutsch 42) is the most studied cluster of them all. There is a wide
range of possible distance and age pairs for this object: 
1.6 kpc and 30 Myr \citep{kronberger2006};
2.0 kpc and 1.3 Myr \citep{kharchenko2013};
6.7 kpc and 3 Myr \citep{morales2013}; and
6.5 kpc and 2-4 Myr \citep{hanson2010}. The last paper from the list \citep{hanson2010}
describes photometric and spectroscopic observations of this cluster and the
quoted distance is the photometric distance, which the authors claim to be in accordance
with the kinematic distance (observed line-of-sight velocities of the brightest
members of the cluster are large, greater than or similar to 70 $kms^{-1}$ and therefore the stars
are assumed to lie at the tangential point). Teutsch 42 contains a WR star and some O and B
stars, its mass is $(4-6) \times 10^3 M_{\odot}$. This age and a presence of evolved massive
stars makes it a very good candidate for the energy source of the bubble A of the background
system. There is still a discrepancy between measured radial velocities of stars in the cluster and 
the gas associated with this cluster.

There are two pulsars in the area with suitable distances from the Sun: J1929+1844 (the distance of 5.3 kpc)
and J1927+1856 (4.8 kpc), \citet{lorimer2002}. One of them could be
ejected from the cluster Teutsch 42, but the probability depends on the distance
of the cluster, which is not precisely known. It is not 
unusual for neutron stars to be expelled from their parental clusters, sometimes with high velocities 
\citep[see][]{moyanoloyola2013}. Unfortunately, at least one of these pulsars seems to be too old 
for the proposed young age of the cluster (J1929+1844: 8 Myr, about twice the age of Teutsch 42).
The other pulsar, J1927+1856, is younger (2 Myr). But even if none of pulsars and the cluster
are correlated, stars of the Teutsch 42 (or similar cluster) are able to produce enough energy to create the bubble A.

Bubble B is the supernova remnant (G54.4-0.3). Its progenitor is unknown.

Bubbles D and E have relatively low energetic demands, compatible with several (one to ten, depending
on the mass) B stars.

\section{Discussion}

\begin{figure}
\centering
\includegraphics[width=9.5cm]{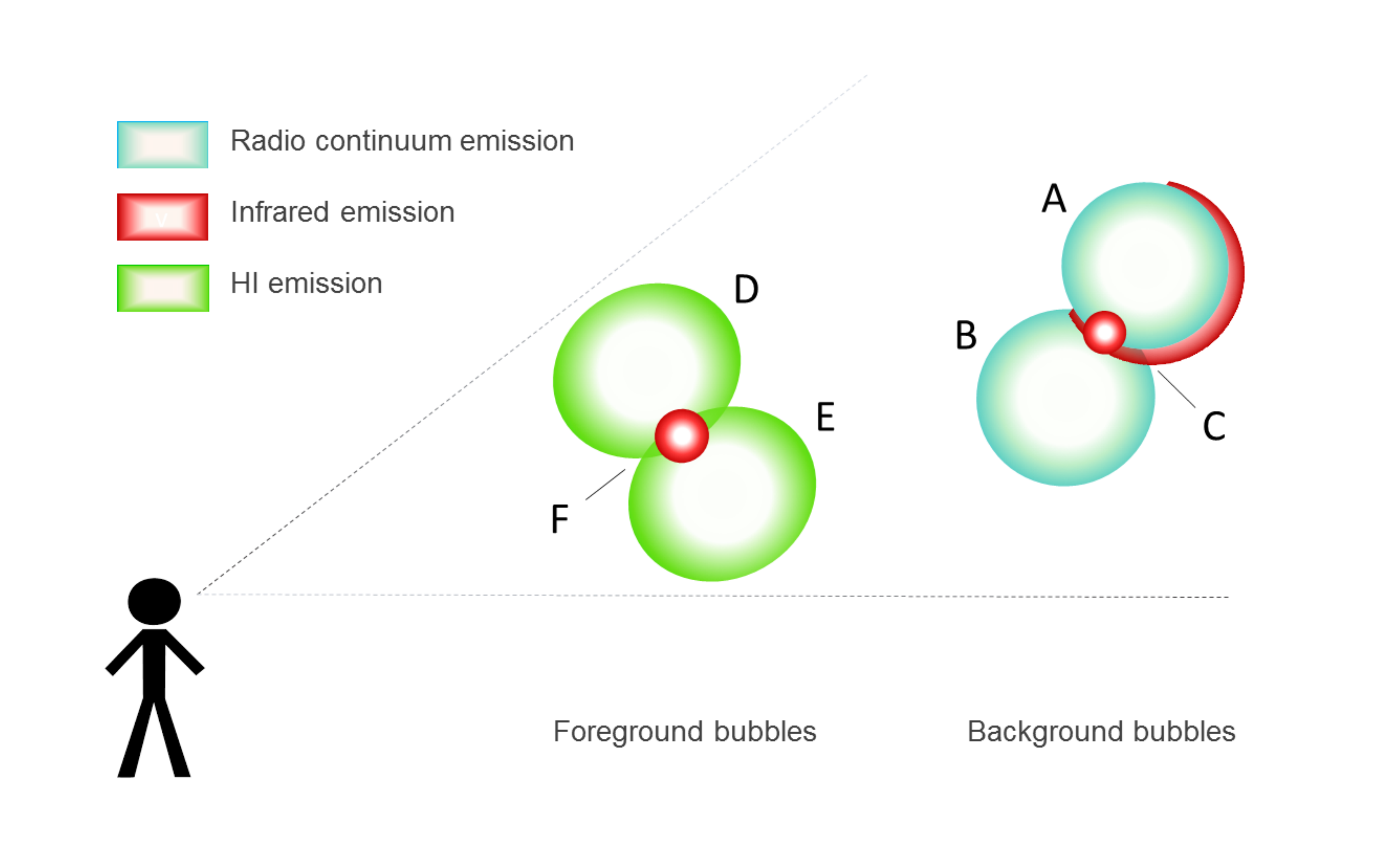}
\caption{A cartoon of the region with all previously known and newly identified bubbles. There are two systems:
a foreground system (larger bubbles D and E and a small bubble F connected to the IR bubble N115);
and the background system (bubbles A = a radio loop GS053.9+0.2, B = SNR G054.4-0.3 and small C = N116+117).}
\label{after}
\end{figure}

Both the foreground and the background system look similar. They are formed by two bubbles with radii
around 20--30 pc and ages of a few million years which collide. A younger and smaller
($\sim 4 \mathrm{pc}$, less than a million years old) bubble
lies at the position of the collision. While this small bubble is best seen in the infrared data, it
is much better visible at all studied wavelengths (IR, radio continuum, HI, CO) than the larger structures. 
Fig. \ref{after} shows our new, more complicated image of the area, where the interacting bubbles
are depicted.

Average densities of the ISM, both corresponding to HI and CO, are higher for the smaller shells. For the bubbles 
with radii around 20 pc, the densities are several $\mbox{H\ cm}^{-3}$ for HI and several 
$10\ \mbox{H}_2\ \mbox{cm}^{-3}$
for CO. For the smaller bubbles the density is (10-20) times higher for HI and 100 times higher for CO.
This is likely the result of the star formation taking place in the densest parts of the ISM, and as the structure 
created by the massive, energy-releasing star, grows, the ISM becomes less and less dense.

These two systems, though they are similar, differ especially in the amount of energy involved
in their creation. The background system, the one in the vicinity of the IR bubble N116+117, is more energetic.
Bubble B is a known supernova remnant \citep{green2009}. Bubble A might also be one: 
our energy estimates are high enough to be consistent with this hypothesis. But it could also be a more energetic 
stellar-wind blown bubble. Other bubbles, i.e. the IR bubbles connected to N116+117 and to N115, and the whole 
foreground system, are probably created by stellar winds or ionizing radiation of massive stars. Energetic 
requirements, 
estimated by the Eq. \ref{eq:etot} of most  of the structures can be fulfilled by one massive star. In the future, 
these bubbles either merge  and create larger structures (if the energy supply is sufficient), 
or they will soon disappear into undetectability and blend with the surrounding. 

There are some young clusters in this area. At least two of them are connected to the
studied bubbles: [BDS2003] 156 to a small bubble C (= N116+117) and [BDS2003] 12 to small F (= N 115).
Teutsch 42 might be the progenitor of the bubble A (= IR loop GS053.9+0.2). The connection of
our studied systems to two young clusters G3CC 69 and 70 and to three young pulsars
(J1927+1856, J1929+1844, J1930+1852) in the area is unclear.

\subsection{Colliding systems}
Perhaps our most interesting finding is that both infrared bubbles (N115 and N1167+117) are associated with collisions of larger
and older bubbles. While we do not claim, that all IR bubbles are created by collisions, these events must be important in
promoting star formation. We therefore propose, that such collision increase the probability of further star formation, 
probably by squeezing the material and increasing its density. 

In addition to the C\&C or RDI scenarios, the collision driven ''Collect \& Collide'' triggered star formation process 
should also be considered.

\begin{acknowledgements}
This study has been supported by the Czech Science Foundation
grant 209/12/1795 and the M\v{S}MT grant LG14013.
\end{acknowledgements}

\bibliographystyle{aa}
\bibliography{colliding_bubbles}

\end{document}